\documentclass[twocolumn, a4paper, aps, prl, 10pt]{revtex4-1}

  \usepackage[utf8]{inputenc}                  
  \usepackage[english]{babel}                  
  \usepackage[T1]{fontenc}                     
  \usepackage{lmodern}                         
  \usepackage{graphicx}                        
  \usepackage{floatflt}                        
  \usepackage{xcolor}                          
  \usepackage{csquotes}
  \usepackage{amsmath}
  \usepackage{mathtools} 
  \usepackage{amsfonts}
  \usepackage{amssymb}
  \usepackage{dsfont}
  \usepackage{mathrsfs}
   \usepackage[pdftitle={Weak localization of magnons in chiral magnets}, breaklinks=true, pdfborder={0 0 0}, bookmarksopen=true, colorlinks=true]{hyperref} 
  \usepackage[]{siunitx} 

\newcommand{\abs}[1]{\ensuremath{ \left| #1 \right| }}                    

\newcommand{\del}[0]{\partial}                

\renewcommand{\rho}[0]{\varrho}
\renewcommand{\theta}[0]{\vartheta}
\renewcommand{\phi}[0]{\varphi}

\renewcommand{\vec}[1]{\mathbf{#1}}        

\newcommand{\ba}[1]{\begin{align} #1 \end{align}}

\begin{document}

\title{Weak localization of magnons in chiral magnets}

\author{Martin Evers}
\author{Cord~A. M\"uller}
\author{Ulrich Nowak}
\affiliation{Fachbereich Physik, Universit\"at Konstanz, 78457 Konstanz, Germany}

\date{\today}

\begin{abstract}
We report on the impact of the Dzyaloshinskii-Moriya interaction on the coherent backscattering of spin waves in a disordered magnetic material.
This interaction breaks the inversion symmetry of the spin-wave dispersion relation, such that $\omega_\vec{k} = \omega_{2\vec{K}^\mathrm{I}-\vec{k}} \neq \omega_{-\vec{k}}$, where $\vec{K}^\mathrm{I}$ is related to the Dzyaloshinskii-Moriya vectors.
As a result of numerical investigations we find that the backscattering peak of a wave packet with initial wave vector $\vec{k}_0$ shifts from $-\vec{k}_0$ to $2\vec{K}^\mathrm{I}-\vec{k}_0$, such that the backscattering wave vector and the initial wave vector are in general no longer antiparallel.
The shifted coherence condition is explained by a diagrammatic approach and opens up an avenue to measure sign and magnitude of the Dzyaloshinskii-Moriya interaction in weakly disordered chiral magnets.   
\end{abstract}

\maketitle

Spin waves or magnons, low-energy excitations of the magnetic groundstate of a solid, have been studied extensively since their first proposal \cite{Kruglyak10_MagnonicsReview,Demokritov13_Magnonics}.
In recent years, they have drawn much attention in connection with new effects such as the spin Seebeck effect \cite{Uchida10_SpinSeebeckInsulatorTransversal},
room-temperature Bose-Einstein condensation \cite{Demokritov06_BEC_Magnons},
magnonic supercurrents \cite{Bozhko16_MagnonSupercurrent} or magnonic topological insulators \cite{Zhang13_MagnonTopologicalInsulator}.
Furthermore, mag\-nonic transport is a promising candidate for future data-processing devices \cite{Serga10_YIG_Magnonics}, because---in contrast to conventional electronic- or spintronic-based technology---magnons do not suffer from Joule heating \cite{Bauer12_SpinCaloritronics}.

Interesting properties arise in chiral magnets,
where the antisymmetric exchange interaction---the Dzya\-lo\-shinskii-Moriya (DM) interaction \cite{Dzyaloshinsky58_WeakFerromagnetismOfAntiferromagnets, Moriya60_AnisotropicSuperexchange_DMI}---induces for example non-colinear ground states \cite{Togawa12_HelicalMagnets_SolitonLattice}, skyrmions \cite{Roessler06_MagneticSkyrmions},
Berry phase materials \cite{Zhang13_MagnonTopologicalInsulator} and a non-inversion symmetric dispersion of the magnons \cite{Udvardi09_SpinWaveSpectraDMI,Moon13_SpinWavePropagation_DMI}.
The DM interaction originates from spin-orbit interactions and is, hence, naturally linked to the upcoming field of spin-orbitronics \cite{Manchon08_NonEquIntrinsicSTT,Manchon09_TheorySpinOrbitTorque,Mellnik14_STT_TopIns,Gomonay16_AFM_DomainWallMotion_SOT,Shiino16_AFM_DomainWallMotion_SOT,Hellman17_InterfacePhenomena}.
In this context it is important to examine the role of defects, because, on the one hand, real magnetic materials inevitably contain some amount of disorder whose impact on device functionality needs to be evaluated. On the other hand, disorder also entails unique effects of its own that may be harnessed for specific applications.

Most prominent in this context is Anderson lo\-cali\-zation \cite{Anderson58_AndLoc} in strongly disordered materials, where coherent transport of waves comes to a complete stop.
But already moderately disordered materials can show interesting weak-localization phenomena, for example the well-known coherent backscattering (CBS) effect \cite{Akkermans07_MesoscopicPhys}.
When a monochromatic wave is launched with wave vector $\vec{k}_0$ into the disordered system,  
CBS can be observed as an 
enhanced average intensity above the incoherent background, usually around the wave vector $-\vec{k}_0$, and thus provides a distinctive measure of phase coherence surviving the ensemble average. 

Recently, we have investigated localization effects in one- and two-dimensional magnetic model systems \cite{Evers15_SpinWaveLoc}. 
It is the purpose of this work to study CBS as a precursor for Anderson localization in chiral magnetic systems, where the presence of the DM interaction leads to a dispersion relation $\omega_\vec{k}$ with broken inversion symmetry, i.e. 
$\omega_\vec{k} = \omega_{2\vec{K}^\mathrm{I}-\vec{k}}\neq \omega_{-\vec{k}}$ ($\vec{K}^\mathrm{I}$ is determined by the DM vectors and is explained below).
In such a system $-\vec{k}_0$ is in general no longer a possible scattering vector under elastic scattering, and one should expect the CBS effect to be weakened, if not entirely suppressed. 
Surprisingly, we find by numerical investigations of an atomistic spin model that CBS survives in such a system with its peak position shifted to $2\vec{K}^\mathrm{I}-\vec{k}_0$.
Remarkably, the height of the CBS peak is not affected at all, in contrast to other model systems \cite{Lenke00_CBS_Faraday} where a shifted coherence condition is generally accompanied by a loss of contrast \cite{Lenke00_CBS_Faraday}.
We will show below that this observation can be explained within a diagrammatic Green's functions approach.

We consider a classical atomistic spin model \cite{Nowak07_SpinModels}, where normalized magnetic moments $\vec{S}^l = \boldsymbol{\mu}^l/\mu_\mathrm{S}$, $l = 1,...,N$, are placed on regular lattice sites $\vec{r}^l$, with $\boldsymbol{\mu}^l$ the magnetic moment of the atom at position $\vec{r}^l$ and $\mu_\mathrm{S}$ its absolute value.
In $d$ dimensions each spin is coupled to its $2d$ nearest neighbors.
The interaction of nearest neighbors $\vec{S}^n$ and $\vec{S}^m$ splits into the isotropic Heisenberg exchange interaction with exchange constant $J > 0$ and the DM interaction, quantified by the DM vectors $\vec{D}^{nm}$ and originating from spin-orbit coupling \cite{Udvardi03_FirstPrinciplesRelativisticSpinWaves}.
In addition we take into account an easy-axis anisotropy in $x$ direction with anisotropy constant $d_x > 0$.
Finally, we also include an external, random magnetic field $\vec{B}(\vec{r}^l) = \vec{B}^l$ that models local disorder.
The Hamiltonian of such a system is
\ba{
  H = & -\frac{J}{2}\sum_{\langle n,m \rangle}\vec{S}^n\cdot\vec{S}^m 
        -\frac{1}{2}\sum_{\langle n,m \rangle}\vec{D}^{nm}\cdot\left(\vec{S}^n\times\vec{S}^m\right) \nonumber\\
      & -\sum_{n}d_x\left(S^n_x\right)^2 -\mu_\mathrm{S}\sum_{n}\vec{B}^n\cdot\vec{S}^n.
}
The spin dynamics in the limit of vanishing damping is governed by the Landau-Lifshitz equation 
\ba{
  \frac{\del \vec{S}^l}{\del t} = -\frac{\gamma}{\mu_\mathrm{S}}\vec{S}^l\times\vec{H}^l, \qquad \vec{H}^l = -\frac{\del H}{\del \vec{S}^l},
  \label{eq:LL}
}
describing the precession of each spin $\vec{S}^l$ in its effective magnetic field $\vec{H}^l$, where $\gamma$ is the gyromagnetic ratio.
It is natural to use $t_J = \mu_\mathrm{S}/\gamma J$ and $B_J = J/\mu_\mathrm{S}$ as units for time and magnetic field, respectively. 

In the case of weak DM interaction, $\abs{\vec{D}^{nm}} < 0.1\,J$, the ground state of the clean system ($\vec{B}^l = 0$) is a ferromagnet parallel to the $x$-axis.
Collective excitations of this ferromagnetic ground state, called magnons or spin waves, can be described by a complex spin-wave amplitude  $\mathcal{S}^l = S_y^l - iS_z^l$ in real space.
In the context of this work, it is more advantageous to describe spin waves via their momentum-space amplitude $\mathcal{S}_\vec{k} = \frac{1}{\sqrt{N}}\sum_n e^{-i \vec{k}\cdot\vec{r}^n} \mathcal{S}^n$.
The dispersion of these spin waves in the linearized limit of small deviations from the clean ground state reads \cite{Udvardi09_SpinWaveSpectraDMI}
\ba{
  \omega_\vec{k} = \frac{1}{t_J}\Bigg[\frac{d_x}{J} & + 2 \sum_{p=1}^{d} \left[1 - \cos(\vec{k}\cdot\vec{a}^p) 
                                                                        +\frac{D_x^p}{J} \sin (\vec{k}\cdot\vec{a}^p) \right]\Bigg]. 
}
The sum runs over the $d$ lattice vectors $\vec{a}^p$, and $\vec{D}^p$ denotes the DM vector $\vec{D}^{nm}$ between two spins $\vec{S}^n$ and $\vec{S}^m$ that are separated by $\vec{a}^p=\vec{r}^m - \vec{r}^n$. Because the ground state is aligned with the easy $x$-axis, the dispersion only depends on the $x$-component of the DM vectors.  
Importantly, the DM interaction breaks the inversion symmetry of the dispersion, $\omega_\vec{k}\neq\omega_{-\vec{k}}$. The sine term in the dispersion shifts the lines of constant frequency, resulting in a dispersion 
\ba{
  \omega_\vec{k} = \omega_{2\vec{K}^\mathrm{I} - \vec{k}} \label{eq:inv_sym}
}
that is instead symmetric with respect to a shifted center of inversion $\vec{K}^\mathrm{I}\neq 0$ determined by  
\ba{
   \vec{K}^\mathrm{I}\cdot\vec{a}^p = -\arctan\left(\frac{D^p_x}{J}\right), \quad p = 1,...,d. \label{KI_from_D}
}
Even in a weakly disordered magnetic material, plane waves are no longer eigenmodes and will be scattered elastically by static, quenched disorder into other accessible modes.
In an inversion-symmetric setting, the CBS signal of a plane wave launched with wave vector $\vec{k}_0$ is found at $-\vec{k}_0$ \cite{Akkermans07_MesoscopicPhys}.  
In case of a non-inversion symmetric dispersion the initial wave cannot be scattered into the $-\vec{k}_0$ state, leading to the question whether CBS can survive in a chiral magnet at all.

\begin{figure*}[t]
  \includegraphics[width=\textwidth]{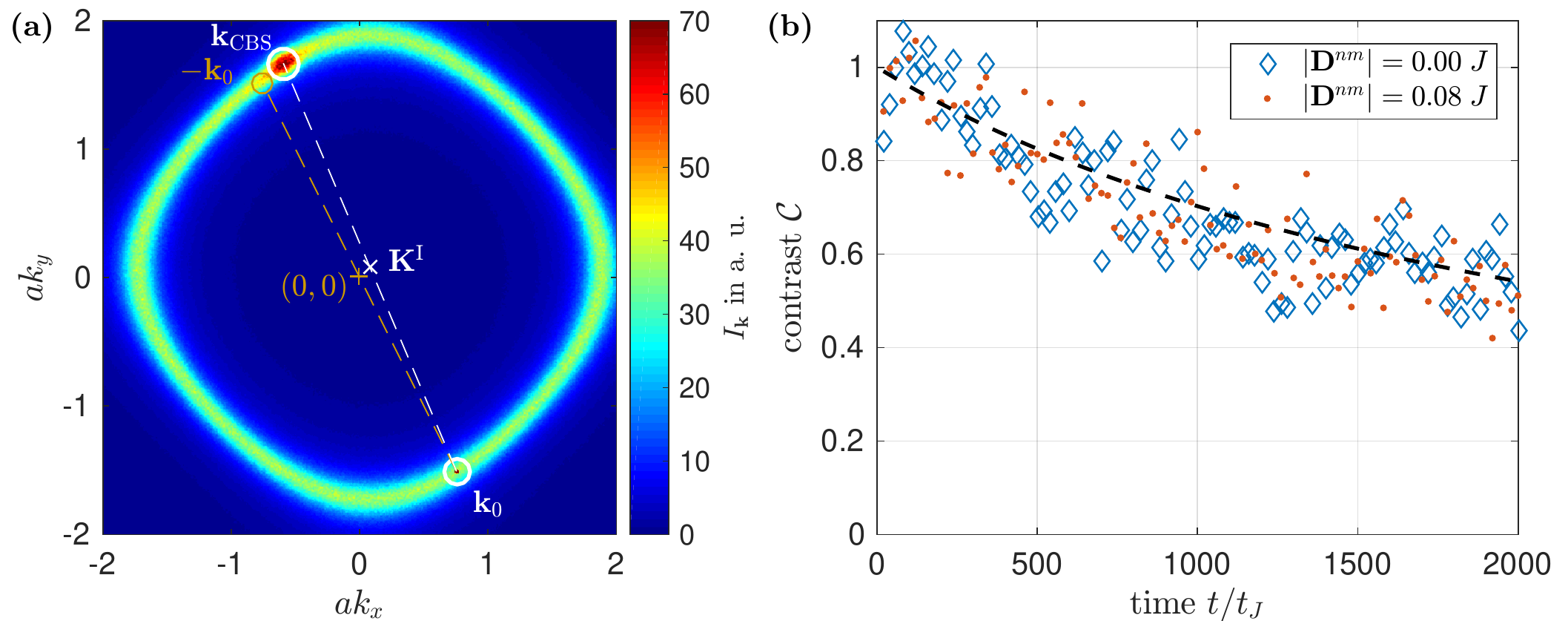}
  \caption{CBS of spin waves in chiral magnets with Dzyaloshinskii-Moriya (DM) interaction.
    (a) Ensemble-averaged spin-wave intensity in $k$-space, $I_\vec{k} = \langle\abs{\mathcal{S}_\vec{k}}^2\rangle$ at time $t = 20\,t_J$. 
    The initial wave packet [Eq.~\eqref{eq:initCond}] is centered at $\vec{k}_0$. 
    A CBS peak rises at $\vec{k}_\mathrm{CBS} \neq -\vec{k}_0$  over the incoherent background.
    Both the center of the Brillouin zone $\Gamma = (0,0)$ and the center of inversion $\vec{K}^\mathrm{I}$ are plotted. 
    The CBS peak position is the conjugate of $\vec{k}_0$ with respect to $\vec{K}^\mathrm{I}$, namely $\vec{k}_\text{CBS}-\vec{K}^\mathrm{I}=\vec{K}^\mathrm{I} - \vec{k}_0$. 
    (b) Time evolution of the CBS contrast with and without DM interaction. The dashed curve is the expected decay $\mathcal{C}(t) = (1 + D t/2\sigma_0^2)^{-1}$ with spin-wave diffusion constant $D\approx 19\,a^2/t_J$ extracted from the real-space diffusive spread of the wave packet.
    Within the noise of the data, there is no observable difference between the cases with and without DM interaction.}
  \label{fig:CBSwithDMI}
\end{figure*}

As a model for thin magnetic films we choose $d = 2$ and a square lattice with lattice constant $a = |\vec{a}^p|$, and perform numerical simulations of spin waves by integrating Eq.~\eqref{eq:LL} using the classical Runge-Kutta method.
The initial condition is a quasi-monochromatic wave packet,
\ba{
  \mathcal{S}^l(t=0) = A \exp\left[ i\vec{k}_0\cdot\vec{r}^l - \left(\vec{r}^l-\vec{r}_0\right)^2/2\sigma_0^2 \right], \label{eq:initCond}
}
with amplitude $A$ and width $\sigma_0$ around the initial position $\vec{r}_0$.
Throughout the paper we use $A = 0.01$, $\sigma_0 = 150\,a$ together with the initial wave vector $\vec{k}_0 = (0.24,-0.48)\pi/a$, except where noted otherwise.  

For concreteness, we consider disorder induced by a longitudinal field $\vec{B}^j = (B,0,0)$ that tries to pin the ferromagnetic orientation at randomly chosen lattice sites $\vec{r}^j$, with $\vec{B}^l = 0$ elsewhere. 
These defect sites are static, uncorrelated, and uniformly distributed with density $\rho$.
In the following, all simulations use $\rho = 0.1$ and $B = 5\,B_J$, followed by an ensemble average $\langle...\rangle$ over 500 defect configurations.

In a first step we choose $D^p_x = -0.08\,J$ and $D^p_{y} = D^p_{z} = 0$.
Following the time evolution, we observe that the average spin-wave intensity in momentum space, $I_\vec{k} = \big\langle\abs{\mathcal{S}_\vec{k}}^2\big\rangle$, 
that is initially concentrated at $\vec{k}_0$ redistributes over the other accessible modes $\vec{k}$ on the energy shell  $\omega_\vec{k}=\omega_{\vec{k}_0}$, slightly broadened by disorder.  
Phase-incoherent diffusion alone would result in a homogeneous distribution, reached on a rather fast time scale given by the transport time \cite{Plisson2013}. Distinctive features above this incoherent background are signatures of phase-coherent processes \cite{Cherroret2012,Evers15_SpinWaveLoc}.
After a few transport times, the spin-wave intensity in $k$-space can thus be written $I_\vec{k}(t) = I^{\mathrm{ic}}_\vec{k} + I^{\mathrm{c}}_\vec{k}(t)$, where $I^{\mathrm{ic}}_\vec{k}$ denotes the incoherent, stationary contribution and $I^{\mathrm{c}}_\vec{k}(t)$ is the part that originates from coherent processes and evolves on longer time scales. 

Figure~\ref{fig:CBSwithDMI}(a) shows $I_\vec{k}(t)$ at a time $t = 20\,t_J$. The incoherent background $I^{\mathrm{ic}}_\vec{k}$ maps out the disorder-broadened energy shell. 
A clear backscattering peak is observable roughly opposite the initial wave vector $\vec{k}_0$. 
Interestingly, the CBS peak position differs from the exact backscattering direction $-\vec{k}_0$ that is well known from the inversion-symmetric setting \cite{Akkermans07_MesoscopicPhys}. 
In the present setting, the backscattering peak appears at the conjugate of $\vec{k}_0$ with respect to the center of inversion $\vec{K}^\mathrm{I}$, namely at $\vec{k}_\mathrm{CBS} = 2\vec{K}^\mathrm{I} - \vec{k}_0$. 
This is remarkable since the backscattering wave vector is not antiparallel to the initial wave vector anymore, but clearly compatible with the symmetry, Eq.~\eqref{eq:inv_sym}, of the dispersion relation.

While the DM interaction is apparently compatible with CBS at early times, though with a shifted peak position, it could very well induce a slight dephasing on longer times scales and therefore result in a faster decay of the CBS peak contrast. 
We investigate this question by recording the time evolution of the CBS contrast $\mathcal{C}(t) = {I^{\mathrm{c}}_{\vec{k}_\mathrm{CBS}}}(t) / {I^{\mathrm{ic}}_{\vec{k}_\mathrm{CBS}}}$ and comparing the cases with and without DM interaction under otherwise identical conditions. 
Even without any additional dephasing processes, the CBS contrast decreases over time because the diffusive CBS interference kernel, whose $k$-space resolution increases over time, is convolved by the finite-width wave packet, Eq.~\eqref{eq:initCond}, leading to an expected decay as $\mathcal{C}(t) = (1 + D t/2\sigma_0^2)^{-1}$, where $D$ is the spin-wave diffusion constant \cite{Cherroret2012,Evers15_SpinWaveLoc}.
The numerical results, shown in figure~\ref{fig:CBSwithDMI}(b), indicate that within the noise of the data the DM interaction does not accelerate this decay significantly and thus does not act as an additional source of dephasing.  

According to linear spin-wave theory within the pres\-ent geometry only the $x$ component of the DM vectors influences the dispersion relation, Eq.~\eqref{eq:inv_sym}.
We have also confirmed this prediction numerically by simulating a system with $D^p_y = D^p_z \neq 0$ and $D^p_x = 0$.
The result is then the same as in the case with $\vec{D}^p = 0$. Because linear spin-wave theory can only be applied for small amplitudes, we have also tested a larger amplitude, $A = 0.2$, and compared again the two cases, $\vec{D}^p = 0$ and $D^p_x = -0.08\,J$.
The decay of the CBS contrast is in both cases much faster than in the linear regime, as a consequence of the non-linearities in the equations of motion that arise for larger amplitudes \cite{Evers15_SpinWaveLoc}.
Still, however, the system with DM interaction shows a decay of the CBS peak just as fast as the system without DM interaction.
We infer from the numerical evidence that the DM interaction does not lower the CBS contrast nor does it lead to faster dephasing, it simply shifts the backscattering wave vector to a different position.

\begin{figure}[t]
  \includegraphics[width=0.9\linewidth]{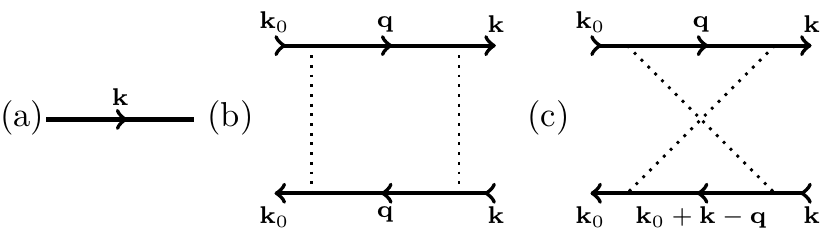}
  \caption{(a) Diagrammatic representation for the ensemble-averaged Green's function, Eq.~\eqref{eq:G_q_omega}.
  (b) Double-scattering contribution to the incoherent stationary intensity kernel, Eq.~\eqref{eq:ladder}. In the present case of isotropic point scatterers it is independent of the external momenta $\vec{k}$ and $\vec{k}_0$.   
  (c) Crossed diagram of the same order, Eq.~\eqref{eq:crossed}, describing the CBS interference contribution. Upon choosing $\vec{k}+\vec{k}_0=2\vec{K}^\text{I}$, this contribution equals the background of (b) and thus yields perfect interference contrast at the shifted position $\vec{k}_\text{CBS}=2\vec{K}^\text{I}-\vec{k}_0$. 
  } 
  \label{fig:diags}
\end{figure}

Our numerical findings in the linear regime can be readily understood via a diagrammatic Green's function approach. 
The rather elementary argument relies, besides the symmetry Eq.~\eqref{eq:inv_sym} of the dispersion, on the fact that a point-like pinning field results in a completely isotropic scattering intensity, noted $U_{\vec{k}\vec{k}'} = U_0$.
As a consequence, the ensemble-averaged, single-magnon Green's function takes the form 
\begin{equation} \label{eq:G_q_omega}
G_\vec{k}(\omega) = [\omega - \omega_\vec{k} - \Sigma(\omega)]^{-1},
\end{equation}
diagrammatically represented by Fig.~\ref{fig:diags}(a),
with a self-energy $\Sigma(\omega)$ that has no momentum dependence on its own. 

As a first building block for the ensemble-averaged intensity, consider then the contribution of scattering from 2 impurities to the incoherent background and compare it to the coherent contribution of the same order.
The incoherent contribution from double scattering to the stationary $\vec{k}$-space distribution is described by the kernel 
\begin{equation} \label{eq:ladder}
L_2(\vec{k}_0,\vec{k},\omega) = \sum_\vec{q} U_0 |G_\vec{q}(\omega)|^2 U_0.    
\end{equation}   
This kernel can be represented by the ladder diagram of a retarded and an advanced amplitude in Fig.~\ref{fig:diags}(b), describing the co-propagation of an amplitude and its complex conjugate along the exact same path in real space. 

The corresponding coherent contribution, stemming from the interference of 2 amplitudes counter-propa\-gating in real space, is given by 
\begin{equation} \label{eq:crossed} 
C_2(\vec{k}_0,\vec{k},\omega) = \sum_\vec{q} U_0 G_\vec{q}(\omega) G^*_{\vec{k}_0+\vec{k}-\vec{q}}(\omega) U_0,   
\end{equation}   
shown in Fig.~\ref{fig:diags}(c). 
Because the impurity vertex $U_0$ is actually independent on momentum, 
only the internal Green's functions depend on the external momenta $\vec{k}$ and $\vec{k}_0$.
The entire crossed diagram thus becomes strictly equal to the corresponding ladder diagram of Fig.~\ref{fig:diags} for $\vec{k} = 2\vec{K}^\text{I} - \vec{k}_0$ since the symmetry Eq.~\eqref{eq:inv_sym} of the dispersion then guarantees that $G_{2\vec{K}^\text{I}-\vec{q}}(\omega)=G_\vec{q}(\omega)$, which makes Eqs.~\eqref{eq:crossed} and \eqref{eq:ladder} equal. 

This argument generalizes in an elementary manner to higher-order scattering processes with a higher number of internal Green's functions obeying the same symmetry. In the end, perfect contrast is achieved order by order, and thus for the entire CBS signal, at the shifted peak position $\vec{k}_\text{CBS}=2\vec{K}^\text{I}-\vec{k}_0$.

A shift of the backscattering peak is known from transport of light in turbid media in the presence of magneto-optical Faraday rotation \cite{Lenke00_CBS_Faraday}. 
However the situation in Faraday experiments differs from the magnetic system studied here in that the shift of the CBS peak is always accompanied by a loss of contrast.
This dephasing is caused by a random shift of the transverse photon polarization at every scattering event, eventually breaking the reciprocity symmetry that would otherwise preserve the CBS contrast.
In our system reciprocity remains intact since the spin-wave polarization is not constrained by transversality and remains unchanged under scattering by scalar impurities. 

The DM interaction originates from spin-orbit coupling of localized or itinerant electrons \cite{Moriya60_AnisotropicSuperexchange_DMI,Levy80_AnisotropicExchange_RKKY+DMI},
which calls for a comparison to electronic transport where spin-orbit coupling has a great impact on weak localization \cite{Hikami80_SpinOrbitMagnetoresistance}.
For electrons strong spin-orbit coupling leads to so-called weak antilocalization \cite{Bergmann83_WeakLoc}, where the amplitudes interfering for CBS collect a phase difference of $2\pi$.
Because electrons are spin-$\frac{1}{2}$ particles, this phase difference implies a sign change that results in  destructive interference such that the scattered intensity is lower in the backscattering direction. 
Obviously, a $2\pi$ phase difference for our bosonic magnons implies constructive interference, so that no antilocalization would be expected from the start.   

In conclusion we have investigated coherent backscattering in a chiral magnetic system with point defects where the dispersion exhibits a broken inversion symmetry that shifts the CBS peak but preserves its contrast. 
Our numerical findings can be understood in terms of a diagrammatic approach using the shifted symmetry of the dispersion, $\omega_\vec{k}=\omega_{2\vec{K}^\mathrm{I} - \vec{k}}$.
The main message is that DM interaction shifts the CBS peak away from the normal $-\vec{k}_0$ direction without altering the contrast nor the decay time.

Since the shift of the CBS peak is directly proportional to the DM vectors by virtue of Eq.~\eqref{KI_from_D}, measuring the CBS position provides a novel way to determine the strength as well as the sign of the DM interaction.
For this, a few alternative methods are available, like Brillouin light scattering \cite{Di15_BLS_DMI_PtCoNi}, spin polarized scanning tunneling microscopy \cite{Bode07_SpinSpiralAFM}, propagating spin wave spectroscopy \cite{Lee16_ElectricalMeasurementDMI} and domain wall motion \cite{Je13_AsymmDomainWallMotionByDMI,Hrabec14_MeasuringailoringDMIPerpendicularMagnetizedThinFilm}.
However, especially for amorphous materials like CoFeB different methods may lead to different results \cite{Soucaille16_ProbingDMI_DomainWallMotion_BLS} and a clear determination of the strength and sign of the DM interaction is still a matter of research.
The CBS effect offers yet another method that may help to clarify this issue.

\begin{acknowledgments}
  This work was performed on the computational resource bwUniCluster funded by the Ministry of Science, Research and Arts and the Universities of the State of Ba\-den-W\"urttemberg, Germany, within the frame\-work program bwHPC.
Financial support by the Deutsche Forschungsgemeinschaft (DFG) via the SFB 767 ``Controlled Nanosystems: Interaction and Interfacing to the Macroscale'' and the SPP 1538 ``Spin Caloric Transport'' is gratefully acknowledged. The authors further thank G.~Maret for helpful discussions.
\end{acknowledgments}

\bibliography{Bibliography}

\end{document}